# Quantum Signature of Anisotropic Singularities in Hydrogen Bond Breaking of Water Dimer

## Author/s

Md Rejwan Ali,

Dept of Material Science & Chemical Engineering

Stony Brook University

Stony Brook, New York

USA

*Corresponding author:   Md Rejwan Ali

<alimd.rejwan@stonybrook.edu>

# Abstract

*Context:*

In all standard force field-based simulations of organic molecules and all types of bio-molecules, polymers; the torsion angle based Hamiltonian term has been an indispensable term to set up molecular simulation tasks. Torsion often termed as a dihedral angle, the coordinate assumes a continuous molecular geometry and energy changes for an angle range from 0 to 360 degrees. However, quantum mechanics-based results presented earlier and in this report show electronic energy will have singularities due to molecular geometry criticality, and torsion based electrotonic energy is not a smooth function due to bond-breaking geometry. In other words contrast to molecular mechanics based results of geometrical continuity, atoms constituting molecules with no stearic clashes, a continuum of geometry under torsion is not feasible as per quantum mechanics particularly when bonding interaction is weak. These results will be novel and significant for force field refinements and in general to validate stable geometry of weak bonded interactions in organic, bio-molecules, and polymers.

*Methods:*

Previously via three different semi-empirical methods, we reported quantum singularities of molecular electronic energies as signature of a chemical bond break-up process in Rivastigmine drug molecule with torsion angle variation around -CO- bond revealed broken chemical moieties of Rivastigmine under experimental X-ray structure identification. In this present work, applying the first principle methods of Hartree-Fock, Density Functional as well as Moller-Plesset techniques, we have reconfirmed the previous general predictions of singularities in molecular electronic energy with torsion angle variation around weak H-bond of water dimer. Due to the quantum nature of the chemical bond breaking process leading to break-point conditions in otherwise connected molecular topology, the singularities in electronic energy are also suggestive of classical large force as needed in bond-breaking process onset. In this current report, we have presented the details of these interesting findings studied on water dimer. These results of quantum singularities can be useful to improve the current bio-molecular force field and to understand reaction chemistry involving bond-breaking under geometrical constraints.

## **Introduction**

In our earlier paper, we have reported for the first time quantum singularities or evidence for quantized energy as a signature of chemical bond break-up process via dihedral angle dependence revealed in the X-ray structure [1-2]. In this present work, via several different ab initio techniques like Hartree-Fock [3], Density Functional [4] as well ,and Moller-Plesset [5] methods and with a higher basis set we have computed the energy of dissociative water dimer as a dihedral angle function and have consistently observed our previous general predictions of singularities in chemical bond breaking process. Due to the quantum nature of chemical bond break-up process leading to break-point conditions in connected molecular topology or in another expression the singularities in water dimer energy are simultaneously also indicative of large classical force at the onset of bond-breaking by torsion of the molecule. Therefore, the results can be used to enhance current molecular simulation techniques. Although purely a quantum mechanical phenomenon, due to weak spectroscopic features, tracking individual chemical bond break-up experimentally in dissociative or other chemical reaction mechanisms can be challenging both in the gas or solution phase by traditional spectroscopies [6]. Spectroscopic detection of such torsion singularities will require advanced state-of-the-art single molecule based force or torque spectroscopy experiments [7-9]. The development of physics-based force field potentials for molecular modeling of bio-macromolecules and large polymer systems had their origin in the first successful molecular dynamics (MD) simulation of liquid argon carried out by implementing the Lennard Jones potential [10-11]. Based on the Born-Oppenheimer theory via assigning an atomic charge at nuclear coordinates with van der Waals radii as soft excluded volume, the MD scheme uses Boltzmann equilibrium assembly properties in correlating ensemble averages with macroscopic properties [12]. In MD simulation force field categories of both bonded and non-bonded terms do connect the atoms by a topology of a network and in practice non-bonded terms practically link the atoms at longer geometrical distances. In molecular simulations, the terms are usually truncated to 10-12 Å in order to save processor time. However, all the Hamiltonian terms in the force field including traditional non-bonded terms are basically a representation of the connectivity of atoms in a bonded network with smooth potential as MD inherently does not simulate any phenomena that is based on electronic structure both for nearest or long distance atomic neighbors. However, as routine methods via using high-performance computing, MD simulations have significant successes in the prediction of X-ray, Cryo-EM structures within reasonable atomic accuracy [13-15]. The connectivity of atoms by molecular mechanics potential reduces accessible phase space particularly to capture critical phenomena when molecular potential might be redefined in the critical transition process; in recent years efforts have been applied in enhanced sampling methods with adaptive polarization and energetics in the assembly [16-17]. Another simulation field used the Monte Carlo (MC) technique mainly based on the pioneering work of Metropolis et al. [18] so far has reported and reproduced some features following the MD trait [19-20]. However, all reported classical MD/MC methods in bio-molecular simulation fields so far have been based on connected molecular topology with distinctive phase space sampling techniques with no categorical difference in molecular mechanics atomic connectivity based on bond breaking or making information by wave mechanics techniques. This is also the case with ab initio MD where refined forces are computed quantum mechanically rather than using molecular mechanics for enhanced phase space sampling but do fall short of capturing actual quantized singularity effects in bond-breaking or bond-making [21].  Due to the above inherent limitations, it is not surprising that normal-mode analysis in MD trajectory analysis does not have the expected success in correlating experimental spectroscopy compared to the ab initio methods [22-23].

    Since its discovery water dimer [24] has been an extensively studied chemical system by various experimental and theoretical techniques over the last few decades [25-28].  It is the simplest form in water cluster formation as has been confirmed by subsequent numerous theoretical and experimental investigations [29-30].  Since its first reporting more than 60 years ago ab initio based studies have been carried out to study and correlate water dimer geometry, dynamics, and spectroscopic features by several methods [31-32]. In this current report, we have repeated and reproduced some of the standard *ab initio* water dimer results previously reported related to optimized geometry and vibrational spectra correlated with experimental methods [33-34]. Finally, results of H-bond breaking as a quantum signature in torsion angle-based electronic energies have been presented as further conclusive evidence of what was previously reported with semi-empirical techniques [1].  In the following sections, we have discussed the methods for computational techniques, discussion, results analysis, and conclusions as well as some quantum computational results with scopes for new experiment design based on reported singularity feature as a logical sequence of the theoretical observations reported in this work.

# Computational Methodologies

## 1. Ground State Geometry of Water Dimer

Ground state geometry of the water dimer has been computed by several first principle techniques as implemented in Gaussian 16 [35] and SPARTAN 18 platform (URL: https://www.wavefun.com; Wavefunction, INC., Irvine, CA). Geometry optimization, bond length, bond angle, HOMO-LUMO orbital, energy gap difference, electrostatic potential map of water dimer have been computed by Moller-Plesset (MP2) with aug-ccPVDz basis, Hartree-Fock (HF) with 6-31G* basis, Density Functional Theory (DFT) at theory level wB97X and B3LYP with 6-31G* basis in gas phase and water medium. DFT with wB97X is a relatively new functional form compared to B3LYP [36] that uses a long-range corrected hybrid density functional with atom-atom damped dispersion correction [37]. Besides DFT, post Hartree-Fock ab initio technique with two types of basis set HF/6-31G* [38-39] and HF/3-21G have been applied for comparison of optimized geometry in the gas phase and water [40-41].

## 2. Water Dimer Binding Energy and Vibrational Spectra

Hydrogen bonds are represented by the well-known 12–10 potential [42] for bond length range between 1.65 to 3.00 Å and acceptor-donor angle 90° < θ < 180°.

$$V(r) = \left(\frac{A}{r^{12}} - \frac{B}{r^{10}}\right) \cos\theta \quad \ldots \ldots \ldots \ldots [1]$$

The theoretical binding energy of water dimer formation was estimated from the hydrogen bond well depth between O4 and H2 (Fig.1) atoms linking water monomers. Fig.2 shows HOMO and LUMO orbital, energy gap, and potential electrostatic map of water dimer computed by B3YLP/6-31G* basis. Likewise, for geometry optimization, several first principle methods have been applied to estimate the binding energy that has reported a value of about 5.2 kcal/mol without zero point energy correction [43-45]. As shown in Table 3, the binding energy value obtained by the MP2 technique is in good consistency with reported values after basis-set superposition corrections [46]. Each energy scan for a specific method has been carried out with the structure initially obtained from geometry optimization. The spectral calculation procedure as mentioned in the above section and discussed below used the same first principle technique. The scan was performed with O4-H2 distance as a variable at 0.1 Å resolution ranging from 1.5 Å to 10 Å as shown in Figs 3 and 4. The dissociation energy has been estimated from the well depth from global minima and the asymptotic unbounded energy threshold computed from the potential plot for each technique. The above method was repeated with O1-O4 [47-48] interaction coordinates as well to compare dimer binding energy results from the H-bonded O4-H2 separation distance method. The results are shown in Table. 3.

For vibrational spectral studies, several methods with different basis sets have been used for bench-marking and best reproducibility of reported theoretical and experimental spectra [33, 49-50]. Infrared vibration peaks were assigned to corresponding different vibrational modes of water dimer atomic vibration groups [51]. Computational vibrational spectra computed by Gaussian 16 have also been reported in the supplementary section. For Spartan 18 based calculations, the systematic error primarily due to the harmonic approximation is corrected by uniformly scaling the amplitude, and spectra lines are broadened to account for finite temperature due primarily to the rotational structure.

## 3. <u>Torsion-Dependent H-bond Energy by MM and QM</u>

Energy for dissociative water dimer molecule around the O4-H2 bond as torsion angle (Figs 9) has been calculated by three separate ab initio methods. A set of three representative ab initio techniques of Hartree-Fock (basis set HF/6-31G*), Moller Plesset (basis MP2/aug-ccPVDZ) and Density Functional Theory (basis set B3YLP/6-31G*) have been applied to map out the torsion-based electronic energy in the entire conformation span for O4-H2 hydrogen bonded atoms as well as O4-O1 interaction atoms forming reaction co-ordinates also shown in Figs 1 and

8. Classical dihedral energy is given by the following equation (symbols have their standard meanings).

$$V(f) = V_N[1 + \cos(N * f - f_o)] \ldots \ldots \ldots \ldots [2]$$

Prior to the energy scan, each structure was geometry optimized via the corresponding technique for ground state parameters and vibrational spectra as reported below. Dihedral energy scans with two different torsion planes of H5O4H2O1 (O4-H2 hydrogen bonded co-ordinates) and H5O4O1H3 (O4-O1 interaction co-ordinates) have been carried out for the entire conformational span of 0° to 360° with 1° resolution. For comparison of the energies with the molecular mechanics-based method has been presented; in fact, the process for both cases was repeated with the Merck Force Field popularly known as MMFF [52], and SYBYL force field [53].

## 4. **Anisotropic Singularities in Torsion Dependent H-bond Electronic Energy Surface**

Anisotropic singularities feature of molecular electronic energy due to weak H-bond of water dimer has been explored both in gas phase and in water medium by semi-empirical AM1 and PM3 techniques which previously we reported can record quantum chemical signature of bond break-up mechanisms. As seen in Fig 9, with fixed O4-H2 or O4-O1 lengths and varied torsion angles for the entire torsional space 0° to 360°, the nature of electronic energetics as a function of torsion angle of water dimer can be explored in the critical range where H-bond interaction should be very weak to hold two water monomers as an integral chemical entity. Since both O1 and O4 atoms are heavier compared to the four hydrogen atoms (H2, H3, H5, and H6), although not bonded by Lewis's definition the interaction distance parameters O1-O4 can serve in good approximation as the center-of-mass distance between two water monomers as has been pointed out in respect to assessing water dimer binding energy. Any singularities in the energetics of such reaction co-ordinates in ($r$, $f$) do imply the break-up of molecular mechanics based connected topology of atoms in otherwise chemically stable molecules. For a fixed $r$, for each case of O4-H2 or O1-O4 atomic distance; torsion angle f has been scanned for the entire conformational space from 0º to 360º at 1º resolution. For O1-O4 the calculations have been extended to water medium both by AM1 and PM3 methods [54-56]. We have explored the electronic energy surface in the critical bond distance range of 2.4 to 2.8 Å. Surface energy plots were generated by converting cylindrical symmetry data (bond length $r$, torsion angle f, and energy) to Cartesian format as per GNUplot 3D graphing routine (URL link: http://www.gnuplot.info/). The results are shown in Figs. 17 and 18 and also in Supplementary Figs 4 and 5.

## Results & Discussions

### 1. Optimized Water Dimer Geometry in Gas Phase and Water Medium

As presented in Tables 1,2, 3, prior to vibrational spectra calculations, water dimer geometry was optimized by several ab initio methods in the gas phase and water medium. The results for geometric parameters of bond length and bond angle in gas phase are shown in Table 1A and 1B. For each method, subsequent vibrational spectra computations have been performed with the optimized geometry as the starting structure as discussed in the next section. As observed below for optimized geometry parameters, all three applied techniques of DFT, MP2, and HF with a basis set of 6-31G*, 6-311+G**, and 6-311+G(2df, 2p) are in general in good consistency as carried out under Spartan 18 software tools with earlier reported works [26, 57]. As in Table 1A and 1B parameters computed by two different *ab initio* tools are reproducible by similar theory level and basis set and are in good agreement. The optimized geometry has been also assessed in the water medium shown in Table 2. Conductor-like Polarizable Continuum Model (CPCM) [58-59] with discretized boundary [60] as implemented in the Spartan 18 package has been applied for water medium computations. Compared to the gas phase, bond distances show a slight reduction in the liquid phase which is consistent with higher packing density in liquid [61].

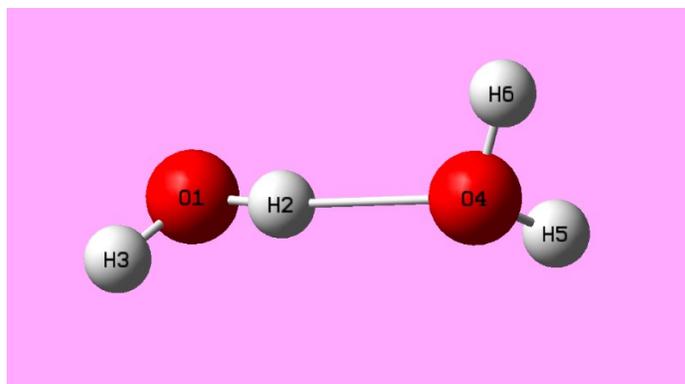

**Fig.1. Water dimer geometry optimized structure and atomic labeling**

**Table. 1A. Optimized geometry of water dimer in gas phase (Atoms are defined in Fig.1)**

| Theory level and basis | O1-O4 distance (Å) | Dimeric H-bond distance of O4H2 (Å) | H2-O1-H3 angle (degree) | H5-O4-H6 angle (degree) | O1-H2-O4 angle (degree) |
|---|---|---|---|---|---|
| HF/STO-3G | 2.771 | 1.762 | 100.78 | 83.84 | 175.75 |
| HF/3-21G | 2.796 | 1.825 | 107.87 | 108.74 | 175.84 |
| HF/6-31G* | 2.991 | 2.068 | 104.62 | 106.24 | 163.32 |
| HF/6-311G* | 2.911 | 1.969 | 107.44 | 108.04 | 177.32 |
| HF/6-311+G** | 3.000 | 2.054 | 106.08 | 106.67 | 178.03 |
| B3LYP/6-31G* | 2.861 | 1.920 | 104.06 | 103.88 | 161.03 |
| B3LYP/ | 2.921 | 1.960 | 105.58 | 105.66 | 171.29 |

| Method | | | | | |
|---|---|---|---|---|---|
| 6-311+G(2df,2p) | | | | | |
| ωB97X-D/6-31G* | 2.862 | 1.915 | 104.13 | 104.52 | 163.92 |
| MP2/cc-pVDZ | 2.909 | 1.944 | 101.78 | 102.40 | 172.95 |
| MP2/aug-cc-pVDZ | 2.916 | 1.951 | 104.28 | 104.20 | 171.12 |

### Table. 1B. Optimized geometry of water dimer in gas phase (Gaussian 16 calculations) (Atoms are defined in Fig.1)

| Method | O1-O4 distance (Å) | O4-H2 dimeric H-bond distance (Å) | H2-O1-H3 angle (degree) | H5-O4-H6 angle (degree) | O1-H2-O4 angle (degree) |
|---|---|---|---|---|---|
| HF/6-31G | 2.842 | 1.885 | 111.60 | 112.18 | 179.60 |
| HF/6-311+G | 2.833 | 1.870 | 112.78 | 112.71 | 179.12 |
| DFT/B3LP-6-31G | 2.777 | 1.794 | 108.99 | 109.61 | 173.65 |
| DFT/ωB97X-D/6-31G | 2.764 | 1.787 | 109.82 | 110.57 | 174.40 |
| MP2/ cc-pVDZ | 2.908 | 1.943 | 101.86 | 105.72 | 172.67 |
| MP2/aug-cc-pVDZ | 2.916 | 1.951 | 104.20 | 104.28 | 171.12 |

### Table. 2. Optimized geometry of water dimer in water medium (Atoms are defined in Fig.1)

| Theory level and basis | O1-O4 distance(Å) | O4-H2 dimeric H-bond distance (Å) | H2-O1-H3 angle (degree) | H5-O4-H6 angle (degree) | O1-H2-O4 angle (degree) |
|---|---|---|---|---|---|
| HF/STO-3G | 2.675 | 1.685 | 100.91 | 100.49 | 178.00 |
| HF/3-21G | 2.728 | 1.747 | 107.94 | 107.20 | 178.52 |

| Method | | | | | |
|---|---|---|---|---|---|
| HF/6-31G* | | | | | |
| HF/6-311G* | 2.873 | 1.927 | 107.05 | 106.77 | 178.73 |
| HF/6-311+G** | 2.947 | 1.999 | 105.78 | 105.79 | 178.29 |
| B3LYP/6-31G* | 2.861 | 1.920 | 104.06 | 103.88 | 161.03 |
| B3LYP/6-311+G(2df,2p) | 2.921 | 1.960 | 105.58 | 105.66 | 171.29 |
| wB97X-D/6-31G* | 2.862 | 1.915 | 104.13 | 104.52 | 163.92 |
| MP2/aug-cc-pVDZ | 2.916 | 1.951 | 104.28 | 104.20 | 171.12 |

## 2. H-bond Dissociative Energy and Vibrational Spectra

A number of reported works have estimated the binding energy of water dimer taking molecular interaction energy as a function of O1-O4 distance instead of O4-H2 shown in Figs 5 and 6. For water dimer structures as centers of mass of both water monomers are localized to heavier O1 and O4 atoms; the dissociation energy of the structure is associated with O1-O4 distance and hence corresponding estimation of the binding energy. As seen in Table 3, both methods give a similar range of binding energies for water dimers and are within consistent ranges of reported values [62]. Particularly the MP2 technique showed high consistency with previous reported work using a similar method [45, 63]. Due to zero point energy fluctuation, the water dimer experimental binding energy is lower than the above values of around 5.2 kcal/mol as reported in a few recent works [25,64]. Also the table of DFT techniques, both by B3YLP and ωB97X-D, shows higher estimates for binding energy as discussed above. In summary, both O1-O4 and O4-H2 methods yield energy values close to the reported water dimer binding energy. In a later section, we also discussed the role of O1-O4 as independent reaction co-ordinates to identify quantum singularities of breaking-up in chemically bonded molecules as has been observed for direct H-bond link between O4-H2 atoms.

**Table. 3 Estimate of Water Dimer Dissociation Energy Global Minima Depth in Energy Plots**

| Theory level and basis set | Dimer binding energy from O4-H2 co-ordinates (kcal/mol) | Dimer binding energy from O1-O4 co-ordinates (kcal/mol) |
|---|---|---|
| HF/STO-3G | -5.85 | -5.89 |
| HF/6-31G* | -5.51 | -5.72 |
| HF/6-311G* | -6.33 | -6.64 |
| MP2/cc-pVDZ | -5.16 | -5.57 |
| MP2/aug-cc-PVDZ | -5.12 | -5.23 |
| B3YLP/6-31G* | -7.49 | -7.53 |
| ωB97X-D/6-31G* | -7.45 | -7.49 |

The table 4. shows results for vibrational spectra mode calculations of the water dimer. The OH- OH-vibration mode is split into symmetric and asymmetric stretches which have distinctively also split into donor and acceptor modes as observed in Figs 7 & 8. Several first principle techniques with different basis set levels have been applied to compute infrared (IR) spectra. The OH vibration region has donor and acceptor modes for both symmetric [3545 cm$^{-1}$ and 3600 cm$^{-1}$] and asymmetric [3715 cm$^{-1}$ and 3730 cm$^{-1}$] modes [65]. Bending motion has a mode[66] around 600 cm$^{-1}$ due to out-of-plane vibration [67]. The computational results show consistency with reported experimental work [33]. In the supplementary section, additional data with similar methods for vibrational spectra calculations have been presented with Gaussian 16 software.

## Table. 4 IR frequencies and their Assignment to Water Dimer Vibrational Modes

| Technique and applied basis set | Out-of-plane vibration mode, cm^(-1) | Bending modes, cm^(-1) | Symmetric stretch modes, cm^(-1) | Asymmetric stretch modes, cm^(-1) |
|---|---|---|---|---|
| HF/3-21G | 832 | 1793, 1854 | 3728 | 3908, 3962 |
| ωB97X-D/6-31G* | 651 | 1620, 1652 | 3511, 3598 | 3688, 3711 |
| B3YLP/6-31G* | 656 | 1628, 1657 | 3449, 3545 | 3635, 3653 |
| EDF2/6-31G* | 689 | 1640, 1671 | 3482, 3598 | 3692, 3709 |
| MP2/cc-PVDz | 667 | 1672, 1713 | 3785, 3836 | 3939 |
| MP2/aug-cc/PVDz | 640 | 1624 | 3703 | 3903, 3924 |

3. **Break-up Signature of Weak H-Bond of Water Dimer**
A. *<u>Energy and Force Estimates for Water Dimer Break-up from Energy Singularities as Dihedral Function</u>*

The hydrogen bond linking two water molecules as reported and computed above has a weak binding energy of around 5.2 kcal/mol without zero point energy corrections. The range of energy makes the water dimer structure very unstable under any bond torsion motion. That dissociative bond-breaking tendency makes the water dimer an ideal system to explore the quantum signature in torsion angle based electronic energy computation that we reported earlier with semi-empirical techniques. For the $H_5O_4H_2O_1$ torsion plane in Fig 1 and Fig 9, dihedral rotation does strain the weak $O_4H_2$ hydrogen bond and reaches critical conformation that breaks the water dimer structure eventually. As observed in each case of HF/6-31G*, B3LYP/6-31G*, MP2/aug-cc/PVDZ based calculations in Figs 11, 12, and 13, except for narrow isolated angle ranges of discontinuities, water dimer H-bond has mostly isotropic flat energy feature as revealed under varied torsion angle scans. This common feature in first principle calculations is quite a distinct contrast to molecular mechanics based torsion energy computed by MMFF and SYBYL methods shown in Fig. 10(A&B). In contrast to QM-based computation of energy, the MM energy curve does not show any singularities rather is continuous along with a smooth first derivative in entire the conformation space in contrast to *ab initio* cases. The amount of energy transition of around ~ 5.5 eV is well above the H-bond binding energy of 0.22 eV or 5.2 Kcal/mol. The electronic energy values are clearly indicative of bond-breaking signature under torsion for weak H bonds in water dimer as the jumps have large energetic features associated with dissociation. In order to account for a more reliable description of hydrogen bond, polarization function with basis set B3LYP/6-311+G(3df,2p) has been used to compute torsion angle-based molecular energy for both O1-O4 and O4-H2 molecular axes around respective dihedral planes. The calculations show the reproducibility of the energy feature with the above methods. The results have been presented in Figs 14 and 15. Based on our hypothesis reported earlier [1] (Ali and Mezei 2021) that results for weakly H-bonded dimers are highly reproducible and independent of quantum computational methods and basis set, we have computed torsion-based water-dimer electronic energy with higher basis set of MP2/aug-cc-PVTZ and DFT-based D-97/D2 calculation with 6-311G+(2df, 2p) basis to account for non-bonded weak dispersive interactions [68-69]. The results have been shown in Figs. 20 and 21. That electronic energy jumps are indeed quantum singularities can also be concluded from torsion-dependent dipole moment variation pattern shown in Figs 20 and 21. For energy singularities in each method, the corresponding dipole moment also shows singularity as well as reversal of slope signs. Both features clearly indicate the criticality of electronic energy and molecular geometry under weak H-bond torsion. DFT Reproducibility works implemented by Gaussian 16 shown in Figs. 22 and 23 also have shown water dimer critical electronic energy pattern for torsion around O4-O1 reaction coordinates. The estimates of dihedral energy slopes in bond-breaking region by B3YLP/6-31G+ and wBxd4/6-31G+ methods are respectively $2\times10^{-3}$ and $8.5\times10^{-3}$ Hartree/Bohr that are equivalent to 0.16 and 0.69 nN. As per experimental data from single molecule force spectroscopy, both force values are several orders of magnitude higher than the usual required hydrogen bond or covalent bond breaking force that are ordered in pN [70-71] (http://www.picotwist.com/index.php?content=smb&option=odg).

## B. Anisotropic Singularities in Torsion-dependent Electronic Energy Surface

Figs 17, 18, and 19 show interesting critical phenomena suggestive of H-bond breaking in 3D energy surface plots as a function of bond length, $r$, and torsion angle, $f$ plotted in the range of 2.0 to 2.8 Å for the entire conformational space of 360° using AM1, PM3 semiempirical and also B3LYP/6-311G+** techniques in the gas phase. Due to a weak H-bond linking two water monomers by around 0.22 eV, the water dimer undergoes predicted chemical cleavages under torsion for O4-H2 distance in bond length threshold and beyond in the range of 2.0 to 2.8 Å. Likewise, estimating water dimer binding energy from O1-O4 interaction co-ordinates even in the absence of Lewis's definition of chemical bonding as explained in the earlier section; interaction distance of O1-O4 and their torsion angle has been explored by H5O4O1H3 plane. But as has been pointed out earlier in case of particularly equivalent water-dimer binding energy, both O4 and O1 nuclear co-ordinates theoretically connect two water molecules, H2-O1-H3 and H5-O4-H6 by the centers of mass point feature of O1-O4 co-ordinates. As a result, any energy singularities with O1-O4 reaction co-ordinateare also indicative of water dimer unstable structure and torsion scan carried out besides H5O4H2O1 torsion plane with explicit O4-H2 H-bonded atoms. In contrast to the above

observed singularities, classical MMFF and SYBYL technique shown in Fig. 16 predictably gives smooth varying energy with dihedral angle variation for O1-O4 as putative H2O linking centers. As a generalized case, we have extended the model to a water medium in addition to H-bonded O4-H2 in the gas phase as we just described in this section. Supplementary Figs 4 and 5 show the energy surface as a function of O4-O1 interaction length and torsion angle of the torsion plane applying AM1 and PM3 techniques. As observed in Supplementary Figs 4 and 5, water dimer energy as torsion function, in general, is isotropic and has a large angle range of stability with a distinctive feature of sharp changes in energy near the molecular critical geometry is observed in the gas phase also. In general, for weak bonds that have the tendency to break down under torsion; ab initio based electronic energy cannot have a finite slope over the entire conformation space in the threshold of bondlength and beyond. For weak H-bonded water dimer systems, electronic energy has shown anticipated discontinuities by all ab initio methods applied in conformation space that we reported earlier for Rivastigmine by semi-empirical methods **[1]** (Ali and Mezei 2021).

## **Conclusions**

The scheme of the connected topology of atoms by molecular mechanics scheme, although has inherent limitations to underpin bond-breaking features via quantum mechanics, still has advantages in defining the collection of atoms for a molecular entity from small organic molecules to large macro-molecules, like protein, DNA/RNA, carbohydrates, lipids, polymers, etc as molecular dynamics is still a robust computational method where ab initio method is limited by atom numbers due to computational bottle-neck. The classical force field theory allows explicit bonding between neighboring atoms via ball-spring models and hence a network of linkages; interestingly quantum mechanics does not eliminate any long distance instantaneous interaction between atoms not bonded via Lewis's definition of proximity or bonding [72-73]. In this paper, we have presented some interesting electronic energy results considering O-H and O-O as equivalent linker coordinates for the water dimer. Electronic energy computational results via molecular mechanics and quantum mechanics methods signifying chemical bond breaking in weak H-bonded water dimer in QM methods and basis set. In summary, in this follow-up article, to validate our previous general predictions [1], water dimer has been selectively chosen for computational studies due mainly to the weak nature of the H-bond linking two water monomers. Under torsion, the weak H-bond shows predicted energy singularities via computational chemistry tools in several quantum chemistry protocols. The method described has been quite distinctive from standard electronic structure literature to explore bond-breaking phenomena using computational tools [74]. Significantly, all standard computational quantum mechanical techniques viz. HF, MP2, and DFT as reported in the current work have given consistent results and conclusions thereof regarding singularity features in electronic energy computations. Currently, another chemical system Binol (ChemSpider ID: 11269, Molecular Formula: C20H14O2) is under study and all our initial results have reproduced singularity features in electronic energy computations. We expect to report the Binol results at an upcoming conference and also in a full article in future publication.

Finally, for internal rotation quantized electronic energy comes as a natural solution when a torsion-like potential is plugged into the Schrodinger time independent equation to compute electronic energy [75]; so discreet energy levels are what nature has over continuum dihedral electronic energy spectrum. In most weak H-bonded or VDW dimer systems, energy singularities should be observed at bond length and positively beyond bond length when the system is under torsion. Interestingly for such molecules according to quantum mechanics continuum of geometry is not allowed with torsion. In contrast force field or molecular mechanics based models do not have such restrictions as long as there are no stearic clashes among atoms. Energy singularities in weak H-bond or VDW dimers and associated geometrical criticality are purely due to the quantum nature of bond break-up and it does not have MM or force-field based analogy as all standard QM calculations are supportive of those facts. With a view to communicating faster the results to scientific communities initial draft and subsequent updated versions of the work have been posted in the pre-print servers: [76,77]. Key results of the work have also been presented in an invited communication at the WATOC 2020 conference in Vancouver, Canada (held in 2022 due to the Covid pandemic).

## Funding Statement

No funding source to declare.

## Acknowledgments

MRA expresses thanks to XSEDE Super Computer Research Resources (Allocation # TG-MCB140084) in Texas Advanced Computing Center in Stampede2 (currently decommissioned) Facilities. The author also acknowledges the scientific invitation and general discussion with Dr. Russell Boyd during the WATOC meeting in Vancouver, Canada held between July 03- 08, 2022. Besides MRA also thanks Dr. Tamar Seideman of Northwestern University for suggesting the Binol system (unpublished results) and Philip Klunzinger of Wavefunction, Inc., Irvine, California for useful suggestions. Many stimulating scientific discussions with Dr. Mihaly Mezei of Icahn School of Medicine at Mount Sinai are greatly acknowledged.

## Supplementary Material

As supporting evidence of the conclusions of the work that the above quantum signature is observable for atoms not only linked by explicit Lewis bonding but rather for any suitable reaction co-ordinates of choice for investigating the stability of chemical bonds or connected topology, we have added additional data on electronic energy of water dimer with O1-O4 as reaction co-ordinates using MP2/augCC_PVDZ and B3YLP/6-311G** basis set. Since the quantum signature results of bond breaking are generalized and valid for any physics conditions, we also have extended the water dimer anisotropic singularities via torsion around O1-O4 reaction co-ordinates in the water medium using AM1 and PM3 as also done for the gas phase. Besides for reproducibility and comparison of vibrational spectra and optimized geometry, results via Gaussian 16 have also been included in the supplementary section. Finally results from the reverse dihedral scan around the O4H2 bond via HF/6-31G* and B3YLP/6-31G* have been appended.